\documentclass[a4paper,12pt]{article}

\usepackage[dvips]{graphicx}
\usepackage{float}
\usepackage{amsmath}
\usepackage{amssymb}
\usepackage{amsfonts}
\usepackage{amscd}
\usepackage{latexsym}

\newtheorem{rem}{Remark}[section]
\newtheorem{defi}{Definition}[section]
\newtheorem{theo}{Theorem}[section]
\newtheorem{lem}{Lemma}[section]
\newtheorem{cor}{Corollary}[section]

\renewcommand{\thefootnote}{\fnsymbol{footnote}}

\newif\ifnotes \notestrue

\def\P{{\cal P}}
\def\Q{{\cal Q}}
\def\R{\mathbb{R}}
\def\diam{\mathop{\rm diam}}
\def\bs{\mathbf s}
\def\bz{\mathbf z}
\def\lt{\left}
\def\rt{\right}
\def\Rd{{\R}^d}
\newcommand{\EXP}{\mathbb{E}}

\newcommand{\IND}{\mathbf{1}}
\newcommand{\defeq}{\stackrel{\rm def}{=}}

\def\qed{\hfill$\Box$}              

\newcounter{mnotecount}[section]

\begin{document}

\begin{center}

{\sc \Large Nonparametric sequential prediction of time
series}
\vspace{1cm}

G\'erard BIAU $^{\mbox{\footnotesize a,}}\footnote{Corresponding
author.}$, Kevin BLEAKLEY $^{\mbox{\footnotesize b}}$,\\
 L\'aszl\'o GY\"ORFI $^{\mbox{\footnotesize c}}$ and
Gy\"orgy OTTUCS\'AK $^{\mbox{\footnotesize c}}$

\vspace{0.5cm}

$^{\mbox{\footnotesize a}}$ LSTA \& LPMA\\Universit\'e Pierre et Marie Curie -- Paris VI\\
Bo\^{\i}te 158,   175 rue du Chevaleret\\
75013 Paris, France\\
\smallskip
\textsf{biau@ccr.jussieu.fr}\\
\bigskip
$^{\mbox{\footnotesize b}}$ Institut de Math\'ematiques et de Mod\'elisation de
Montpellier\\ UMR CNRS 5149, Equipe de Probabilit\'es et Statistique\\
Universit\'e Montpellier II, CC 051\\ Place Eug\`ene Bataillon, 34095 Montpellier Cedex 5, France\\
\smallskip
\textsf{bleakley@math.univ-montp2.fr}\\
\bigskip
$^{\mbox{\footnotesize c}}$ Department of Computer Science and Information Theory\\
Budapest University of Technology and Economics\\
H-1117 Magyar Tud\'osok krt. 2, Budapest, Hungary\\
\smallskip
\textsf{\{gyorfi,oti\}@szit.bme.hu}\\

\vspace{0.5cm}

\end{center}


\begin{abstract}
\noindent Time series prediction covers a vast field of every-day statistical
applications in medical, environmental and economic domains.
In this paper we develop nonparametric
prediction strategies based on the combination of a set of ``experts'' and show the
universal consistency of these strategies under a minimum of conditions. We perform an in-depth analysis of real-world data sets
and show that these nonparametric strategies are more flexible, faster and generally outperform
ARMA methods in terms of normalized cumulative prediction error.\\


\end{abstract}

\renewcommand{\thefootnote}{\arabic{footnote}}

\setcounter{footnote}{0}

\section{Introduction}
The problem of time series analysis and prediction has a long and
rich history, probably dating back to the pioneering work of Yule in
1927 \cite{Yule}. The application scope is vast, as time series
modeling is routinely employed across the entire and diverse range
of applied statistics, including problems in genetics, medical
diagnoses, air pollution forecasting, machine condition
monitoring, financial investments, marketing and econometrics. Most
of the research activity until the 1970s was concerned with
parametric approaches to the problem whereby a simple, usually
linear model is fitted to the data (for a comprehensive account we
refer the reader to the monograph of Brockwell and Davies
\cite{Brockwell}). While many appealing mathematical properties of
the parametric paradigm have been established, it has become clear
over the years that the limitations of the approach may be rather
severe, essentially due to overly rigid constraints which are
imposed on the processes. One of the more promising solutions to
overcome this problem has been the extension of classic
nonparametric methods to the time series framework (see for example
Gy\"orfi, H\"ardle, Sarda and Vieu \cite{Laci1} and Bosq \cite{Bosq}
for a review and references).

Interestingly, related schemes have been proposed in the context of sequential investment strategies for financial markets. Sequential investment strategies are allowed to use information about the market collected from the past and determine at the beginning of a training period a portfolio, that is, a way to distribute the current capital among the available assets. Here, the goal of the investor is to maximize their wealth in the long run, without knowing the underlying distribution generating the stock prices. For more information on this subject, we refer the reader to Algoet \cite{Algoet1}, Gy\"orfi and Sch\"afer \cite{Laci5}, Gy\"orfi, Lugosi and Udina \cite{Laci3}, and Gy\"orfi, Udina and Walk \cite{Laci6}.

The present paper is devoted to the nonparametric problem of
sequential prediction of real valued sequences which we
do not require to
necessarily satisfy the classical statistical assumptions for bounded, autoregressive
or Markovian processes. Indeed, our goal is to show powerful
consistency results under a strict minimum of conditions.
To fix
the context, we suppose that at each time instant $n=1, 2, \hdots$,
the statistician (also called the {\it predictor} hereafter) {\it is
asked to guess the next outcome $y_n$} of a sequence of real numbers
$y_1, y_2, \hdots$ with knowledge of the past $y_1^{n-1}=(y_1,
\hdots,y_{n-1})$ (where $y_1^0$ denotes the empty string) and the
side information vectors $x_1^n=(x_1,\ldots,x_n)$, where $x_n \in
\Rd$. In other words, adopting the perspective of on-line learning,
the elements $y_0, y_1, y_2, \hdots$ and $x_1, x_2, \hdots$ are revealed
one at a time, in order, beginning with
$(x_1,y_0),(x_2,y_1),\ldots$, and the predictor's estimate of $y_n$ at time
$n$ is based on the strings $y_1^{n-1}$ and $x_1^{n}$. Formally, the
strategy of the predictor is a sequence $g=\{g_n\}_{n=1}^{\infty}$ of
forecasting functions
\[
   g_n: \lt(\Rd\rt)^n \times \R^{n-1} \to \R
\]
and the prediction formed at time $n$ is just  $g_n(x_1^n,y_1^{n-1})$.

Throughout the paper we will suppose that
$(x_1,y_1),(x_2,y_2),\ldots$  are realizations of random
variables $(X_1,Y_1),(X_2,Y_2),\ldots$ such that the process $\{(X_n,Y_n)\}_{-\infty}^{\infty}$ is jointly stationary and
ergodic. 

After $n$ time instants, the ({\it normalized}) {\it cumulative squared prediction error} on the strings $X_1^n$ and $Y_1^n$ is
$$L_n(g)=\frac{1}{n}\sum_{t=1}^n\left(g_t(X_1^{t},Y_1^{t-1})-Y_t\right)^2.$$
Ideally, the goal is to make $L_n(g)$ small. There is, however, a fundamental limit for the predictability of the sequence, which is determined by a result of Algoet \cite{Algoet2}: for any prediction strategy $g$ and jointly stationary ergodic process $\{(X_n,Y_n)\}_{-\infty}^{\infty}$,
\begin{equation}\liminf_{n \to \infty} L_n(g) \geq L^* \quad \mbox{almost surely},\label{algoet}\end{equation}
where
$$L^*=\mathbb E \left\{\left(Y_0-\mathbb E \left\{ Y_0 |X_{-\infty}^{0},Y_{-\infty}^{-1}\right\}\right)^2\right\}$$
is the minimal mean squared error of any prediction for the value of $Y_0$ based on the infinite past observation sequences $Y_{-\infty}^{-1}=(\hdots, Y_{-2}, Y_{-1})$ and
$X_{-\infty}^{0}=(\hdots, X_{-2}, X_{-1})$. Generally, we cannot hope to design a strategy whose prediction error exactly achieves the lower bound $L^*$. Rather, we require that $L_n(g)$ gets arbitrarily close to $L^*$ as $n$ grows. This gives sense to the following definition:
\begin{defi}
A prediction strategy $g$ is called universally consistent with respect to a class $\,\mathcal C$ of stationary and ergodic processes $\{(X_n,Y_n)\}_{-\infty}^{\infty}$ if for each process in the class,
$$ \lim_{n \to \infty} L_n(g)=L^*\quad \mbox{almost surely}.$$
\end{defi}
Thus, universally consistent strategies asymptotically achieve the
best possible loss for all processes in the class. Algoet
\cite{Algoet1} and Morvai, Yakowitz and Gy\"orfi \cite{Morvai}
proved that there exist universally consistent strategies with
respect to the class $\mathcal C$ of all bounded, stationary and
ergodic processes. However, the prediction algorithms
discussed in these papers are either very complex or have an
unreasonably slow rate of convergence, even for well-behaved
processes. Building on the methodology developed in recent years for
prediction of individual sequences (see Cesa-Bianchi and Lugosi
\cite{Cesa} for a survey and references), Gy\"orfi and Lugosi
introduced in \cite{Laci2} a histogram-based prediction strategy
which is ``simple'' and yet universally consistent with respect to the
class $\mathcal C$. A similar result was also derived independently by
Nobel \cite{Nob03}.
Roughly speaking, both methods consider several
partitioning estimates (called {\it experts} in this context) and
combine them at time $n$ according to their past performance. For
this, a probability distribution on the set of experts is generated,
where a ``good'' expert has relatively large weight, and the average
of all experts' predictions is taken with respect to this
distribution.

The purpose of this paper is to further investigate
nonparametric expert-oriented strategies for unbounded time series
prediction. With this aim in mind, in Section 2.1 we briefly recall the
histogram-based prediction strategy initiated in \cite{Laci2}, which was recently extended to unbounded processes by Gy\"orfi and Ottucs\'ak \cite{Laci4}. In Section 2.2 and 2.3 we offer two ``more
flexible'' strategies, called respectively  {\it kernel} and {\it
nearest neighbor-based} prediction strategies, and state their
universal consistency with respect to the class of all
(non-necessarily bounded) stationary and ergodic processes with
finite fourth moment. In Section 2.4 we consider as an
alternative a prediction strategy based on combining generalized linear
estimates. In Section 2.5 we use the techniques of the previous
section to give a simpler prediction strategy for stationary Gaussian
ergodic processes. Extensive experimental results based on
real-life data sets are discussed in Section 3, and proofs of the
main results are given in Section 4.
\section{Universally consistent prediction strategies}
\subsection{Histogram-based prediction strategy}
In this section, we briefly describe the histogram-based
prediction scheme due to Gy\"orfi and Ottucs\'ak \cite{Laci4} for {\it unbounded} stationary and ergodic sequences. The strategy is
defined at each time instant as a convex combination of {\it
elementary predictors} (the so-called {\it experts}), where the
weighting coefficients depend on the past performance of each
elementary predictor. To be more precise, we first define an
infinite array of experts $h^{(k,\ell)}$, $k, \ell=1,
2, \hdots$ as follows. Let $\P_{\ell}=\{A_{\ell,j},
j=1,2,\ldots,m_{\ell}\}$ be a sequence of finite partitions of $\R^d$,
and let $\Q_{\ell}=\{B_{\ell,j}, j=1,2,\ldots,m'_{\ell}\}$ be a
sequence of finite partitions of $\R$. Introduce the corresponding
quantizers:
\[
F_{\ell}(x)=j, \mbox{ if } x \in A_{\ell,j}
\]
and
\[
G_{\ell}(y)=j, \mbox{ if } y \in B_{\ell,j}.
\]
To lighten notation a bit, for any $n$ and $x_1^n \in
(\R^d)^n$, we write $F_{\ell}(x_1^n)$ for the sequence
$F_{\ell}(x_1),\ldots,F_{\ell}(x_n)$ and similarly, for $y_1^n \in
\R^n$ we write $G_{\ell}(y_1^n)$ for the sequence
$G_{\ell}(y_1),\ldots,G_{\ell}(y_n)$.

The sequence of experts $h^{(k,\ell)}$, $k,\ell =1,2,\ldots$ is defined as follows. Let $J_n^{(k,\ell)}$ be the
locations of the matches of the last seen strings $x^{n}_{n-k}$ of
length $k+1$ and $y^{n-1}_{n-k}$ of length $k$ in the past according
to the quantizer with parameters $k$ and $\ell$:
\[
J_{n}^{(k,\ell)}=\left\{k<t<n:
F_{\ell}(x^{t}_{t-k})=F_\ell(x^{n}_{n-k}),G_{\ell}(y^{t-1}_{t-k})=G_\ell(y^{n-1}_{n-k})\right\},
\]
and introduce the truncation function
\[
 T_a(z) = \left \{ \begin{array}{ll}
       a & \mbox{if $z > a$;} \\
       z & \mbox{if $|z| \le a$;}\\
       -a & \mbox{if $z < -a$.}
                \end{array}   \right.
\]
Now define the elementary predictor
 $h_n^{(k,\ell)}$ by
\[
  h_n^{(k,\ell)} (x_1^{n},y_1^{n-1})
  =  T_{n^{\delta}}\left(\frac{1}{|J_{n}^{(k,\ell)}|}\sum_{\{t \in
  J_{n}^{(k,\ell)}\}}y_t\right), \qquad n>k+1,
\]
where $0/0$ is defined to be $0$ and
\[
0<\delta <1/8.
\]
Here and throughout, for any finite set $J$, the notation $|J|$ stands for the size of $J$. We note that the expert $h_n^{(k,\ell)}$ can be interpreted as a (truncated) histogram regression function estimate drawn in $(\R^d)^{k+1} \times \R^k$ (Gy\"orfi, Kohler, Krzy\.zak and Walk \cite{Laciregressionbook}).

The proposed prediction algorithm proceeds with an exponential
weighting average method. Formally, let $\{q_{k,\ell}\}$ be a
probability distribution on the set of all pairs $(k,\ell)$ of
positive integers such that for all $k$ and $\ell$, $q_{k,\ell}>0$. Fix a learning parameter
$\eta_n>0$, and define the weights
\[
w_{k,\ell,n}=q_{k,\ell}e^{-\eta_n(n-1)L_{n-1}(h^{(k,\ell)})}
\]
and their normalized values
\[
p_{k,\ell,n}=\frac{w_{k,\ell,n}}{\sum_{i,j=1}^{\infty}w_{i,j,n}}.
\]
The prediction strategy $g$ at time $n$ is defined by
\begin{equation*}
  g_n(x_1^n,y_1^{n-1})
  =
  \sum_{k,\ell=1}^{\infty}p_{k,\ell,n}h_n^{(k,\ell)}(x_1^n,y_1^{n-1}),
\qquad n=1,2,\ldots
\end{equation*}

It is proved in \cite{Laci4} that this scheme is universally
consistent with respect to the class of all (non-necessarily bounded) stationary and
ergodic processes with finite fourth moment, as stated in the
following theorem. Here and throughout the document, $\|\cdot\|$ denotes the Euclidean norm.
\begin{theo}[Gy\"orfi and Ottucs\'ak \cite{Laci4}]
\label{cons} Assume that
\begin{itemize}
\item[(a)] The sequence of partitions $\P_{\ell}$ is nested, that
is, any cell of ${\cal P}_{\ell+1}$ is a subset of a cell of ${\cal
P}_{\ell}$, $\ell=1,2,\ldots$;

\item[(b)] The sequence of partitions $\Q_{\ell}$ is nested;

\item[(c)] The sequence of partitions $\P_{\ell}$ is
asymptotically fine, i.e., if \[ \diam(A) = \sup_{x,y \in A} \|x-y\|
\]
denotes the diameter of a set, then for each sphere $S$ centered at
the origin
\[
\lim_{\ell\to \infty}
 \max_{j:A_{\ell,j}\cap S\ne \emptyset}\diam(A_{\ell,j})=0;
\]

\item[(d)] The sequence of partitions $\Q_{\ell}$ is
asymptotically fine.
\end{itemize}
Then, if we choose the learning parameter $\eta_n$ of the algorithm as
$$\eta_n=\frac{1}{\sqrt{n}},$$
the histogram-based prediction scheme $g$ defined above is
universally consistent with respect to the class of all jointly stationary and ergodic
processes such that
\[
\EXP\{Y_0^4\}<\infty.
\]
\end{theo}
The idea of combining a collection of concurrent estimates was
originally developed in a non-stochastic context for on-line
sequential prediction from deterministic sequences (see Cesa-Bianchi
and Lugosi \cite{Cesa} for a comprehensive introduction). Following the terminology of
the prediction literature, the combination of different procedures
is sometimes termed {\it aggregation} in the stochastic context. The
overall goal is always the same: use aggregation to improve
prediction. For a recent review and an updated list of references,
see Bunea and Nobel
\cite{Bunea} and Bunea, Tsybakov and Wegkamp \cite{BTW}.

\subsection{Kernel-based prediction strategies}
We introduce in this section a class of {\it kernel-based} prediction
strategies for (non-necessarily bounded) stationary and ergodic sequences. The main advantage of this approach in contrast to the
histogram-based strategy is that it replaces the rigid discretization of the past appearances by
more flexible rules. This also often leads to faster algorithms in practical applications.

To simplify the notation, we start with the  simple
``moving-window'' scheme, corresponding to a uniform kernel
function, and treat the general case briefly later. Just like
before, we define an array of experts $h^{(k,\ell)}$, where $k$ and
$\ell$ are positive integers. We associate to each pair
$(k,\ell)$ two radii $r_{k,\ell}>0$ and $r'_{k,\ell}>0$ such
that, for any fixed $k$
\begin{equation}
\lim_{\ell \to \infty} r_{k,\ell}=0, \label{eq:radius1}
\end{equation}
and
\begin{equation}
\lim_{\ell \to \infty} r'_{k,\ell}=0. \label{eq:radius2}
\end{equation}
Finally, let the location of the matches be
\[
J_{n}^{(k,\ell)}=\left\{k<t<n : \|x^{t}_{t-k}-x^{n}_{n-k}\| \le
r_{k,\ell},\ \|y^{t-1}_{t-k}-y^{n-1}_{n-k}\|\le r'_{k,\ell}\right\}~.
\]
Then the elementary expert
$h_n^{(k,\ell)}$ at time $n$ is defined by
\begin{equation}\label{troisun}
  h_n^{(k,\ell)} (x_1^{n},y_1^{n-1})
  =  T_{\min\{n^{\delta},\ell\}}\left(\frac{\sum_{\{t \in
  J_{n}^{(k,\ell)}\}}y_t}{|J_{n}^{(k,\ell)}|}\right), \qquad
  n>k+1,
\end{equation}
where $0/0$ is defined to be $0$ and
\[
0<\delta <1/8~.
\]
The pool of experts is mixed the same way as in the case of the
histogram-based strategy. That is, letting $\{q_{k,\ell}\}$ be a
probability distribution  over the set of all pairs $(k,\ell)$ of
positive integers such that $q_{k,\ell} >0$ for all $k$ and $\ell$,
for $\eta_n>0$, we define the weights
$$w_{k,\ell,n}=q_{k,\ell}e^{-\eta_n(n-1)L_{n-1}(h^{(k,\ell)})}$$
together with their normalized values
\begin{equation}
\label{eq:pkln}
p_{k,\ell,n}=\frac{w_{k,\ell,n}}{\sum_{i,j=1}^{\infty}w_{i,j,n}}.
\end{equation}
The general prediction scheme $g_n$ at time $n$ is then defined by
weighting the experts according to their past performance and the
initial distribution $\{q_{k,\ell}\}$:
\begin{equation*}
 g_n(x_1^n,y_1^{n-1})=\sum_{k,\ell=1}^{\infty}p_{k,\ell,n}h_n^{(k,\ell)}(x_1^n,y_1^{n-1}),
\qquad n=1,2,\ldots
\end{equation*}

\begin{theo}
\label{th:kernel}
Denote by $\mathcal C$ the class of all
jointly stationary and ergodic processes
$\{(X_n,Y_n)\}_{-\infty}^{\infty}$ such that $\mathbb E \{Y_0^4\}<\infty$.
Choose the learning parameter $\eta_n$ of the algorithm as
\[
\eta_n=\frac{1}{\sqrt{n}}\,,
\]
and suppose that (\ref{eq:radius1}) and (\ref{eq:radius2}) are verified. Then the moving-window-based prediction strategy defined above is universally consistent with respect
to the class $\mathcal C$.
\end{theo}

\noindent The proof of Theorem \ref{th:kernel} is in Section \ref{proofs}.
This theorem may be extended to a more general class of
kernel-based strategies, as introduced in the next remark.

\begin{rem}[{\sc General kernel function}]
Define a {\it kernel function} as any map $K:\mathbb R_+ \to \mathbb R_+$.
The kernel-based strategy parallels the moving-window scheme defined above, with
the only difference that in definition (\ref{troisun}) of
the elementary strategy, the regression function estimate is
replaced by
\begin{align*}
  & h_n^{(k,\ell)} (x_1^{n},y_1^{n-1})\\
  & \quad =  T_{\min\{n^{\delta},\ell\}}\left(\frac{\sum_{\{k < t < n\}}
  K\left( \|x_{t-k}^{t}-x_{n-k}^{n}\|/r_{k,\ell}\right)
  K\left(\|y_{t-k}^{t-1}-y_{n-k}^{n-1}\|/r'_{k,\ell}\right)y_t}
  {\sum_{\{k < t < n\}} K\left(\|x_{t-k}^{t}-x_{n-k}^{n}\|/r_{k,\ell}\right)
  K\left(\|y_{t-k}^{t-1}-y_{n-k}^{n-1}\|/r'_{k,\ell}\right)} \right).
\end{align*}
Observe that if $K$ is the naive kernel $K(x)=\mathbf 1_{\{x\leq
1\}}$ $($where $\mathbf 1$ denotes the indicator function and $x \in
\mathbb R_+)$, we recover the moving-window strategy discussed
above. Typical nonuniform kernels assign a smaller weight to the
observations $x_{t-k}^{t}$ and $y_{t-k}^{t-1}$ whose distance
from $x_{n-k}^{n}$ and $y_{n-k}^{n-1}$ is larger. Such kernels
promise a better prediction of the local structure of the
conditional distribution.
\end{rem}
\subsection{Nearest neighbor-based prediction strategy}
This strategy is yet more robust with respect to the kernel
strategy and thus also with respect to the histogram strategy. This is because
it does not suffer from the scaling problems of histogram and kernel-based
strategies where the quantizer and the radius have to be carefully
chosen to obtain ``good'' performance. 

To introduce the strategy, we start again by defining an infinite array of experts
$h^{(k,\ell)}$, where $k$ and $\ell$ are positive integers. Just
like before, $k$ is the length of the past observation vectors
being scanned by the elementary expert and, for each
$\ell$, choose $p_{\ell} \in (0,1)$
such that
\begin{equation}
\lim_{\ell \to \infty} p_{\ell}=0 \label{eq:pl}\,,
\end{equation}
and set
$$\bar \ell=\lfloor p_{\ell} n\rfloor$$
(where $\lfloor . \rfloor$ is the floor function).
At time $n$, for fixed $k$ and $\ell$ ($n > k + \bar \ell +1)$, the
expert searches for the $\bar \ell$ nearest neighbors (NN) of the
last seen observation $x_{n-k}^{n}$ and $y_{n-k}^{n-1}$ in the past and 
predicts accordingly. More precisely, let
\begin{eqnarray*}
J_{n}^{(k,\ell)}\!\!
=\!\!\!\!&\big\{\!\!\!\!& k<t<n  : (x_{t-k}^{t},y_{t-k}^{t-1}) \mbox{ is among
the } \bar \ell \mbox{ NN of } (x_{n-k}^{n},y_{n-k}^{n-1}) \mbox{ in } \\
&&(x_1^{k+1},y_1^k), \hdots, (x_{n-k-1}^{n-1}, y_{n-k-1}^{n-2})\big\}
\end{eqnarray*}
and introduce the elementary predictor
$$
h_n^{(k,\ell)}(x_1^{n},y_1^{n-1})=T_{\min\{n^{\delta},\ell\}}\left(\frac{\sum_{\{t
\in J_{n}^{(k,\ell)}\}}y_t}{| J_{n}^{(k,\ell)}|}\right)
$$
if the sum is non void, and $0$ otherwise. Next,  set
\[
0<\delta<\frac{1}{8}.
\]
Finally, the experts are mixed as before: starting from an initial probability distribution $\{q_{k,\ell}\}$, the aggregation scheme is
\begin{equation*}
g_n(x_1^n,y_1^{n-1})=\sum_{k,\ell=1}^{\infty}p_{k,\ell,n}h_n^{(k,\ell)}(x_1^{n},y_1^{n-1}),
\qquad n=1,2, \hdots,
\end{equation*}
where the probabilities $p_{k,\ell,n}$ are the same as in
(\ref{eq:pkln}).

\begin{theo}
\label{th:nn} Denote by $\mathcal C$ the class of all jointly
stationary and ergodic processes $\{(X_n,Y_n)\}_{-\infty}^{\infty}$ such
that $\mathbb E \{Y_0^4\}<\infty$. Choose the parameter $\eta_n$ of the
algorithm as
\[
\eta_n=\frac{1}{\sqrt{n}}\,,
\]
and suppose that (\ref{eq:pl}) is verified. Suppose also that for each vector $\bs$ the random
variable
\[
\|(X_1^{k+1},Y_1^{k})-\bs\|
\]
has a continuous distribution function. Then the nearest neighbor prediction strategy defined above is
universally consistent with respect to the class $\mathcal C$.
\end{theo}
The proof is a combination of the proof of Theorem \ref{th:kernel} and
the technique used in \cite{Laci6}.
\subsection{Generalized linear prediction strategy}
\label{glin}
This section is devoted to an alternative way of defining a universal predictor for stationary and ergodic processes. It is in effect an extension of the approach presented in Gy\"orfi and Lugosi \cite{Laci2} to non-necessarily bounded processes. Once again, we apply the method described in the previous sections to combine elementary predictors, but now we use elementary predictors which are generalized linear predictors. More precisely, we define an infinite array of
elementary experts $h^{(k,\ell)}$, $k,\ell=1,2,\ldots$ as
follows. Let $\{\phi_j^{(k)}\}_{j=1}^{\ell}$ be real-valued
functions defined on ${(\R^d)}^{(k+1)}\times \R^k$. The elementary
predictor $h_n^{(k,\ell)}$ generates a prediction of form
\[
h_n^{(k,\ell)}(x_1^n,y_1^{n-1})
  = T_{\min\{n^{\delta},\ell\}}\left(\sum_{j=1}^{\ell} c_{n,j}\phi_j^{(k)}(x_{n-k}^n,y_{n-k}^{n-1})\right)\,,
\]
where the coefficients $c_{n,j}$ are calculated according to the
past observations $x_1^n$, $ y_1^{n-1}$, and
\[
0<\delta<\frac{1}{8}.
\]
Formally, the coefficients $c_{n,j}$ are defined as the real numbers which minimize the criterion
\begin{equation}\label{eq:cnj}
\sum_{t=k+1}^{n-1} \left(\sum_{j=1}^{\ell}
c_{j}\phi_j^{(k)}(x_{t-k}^t,y_{t-k}^{t-1})-y_t\right)^2
\end{equation}
if $n>k+1$, and the all-zero vector otherwise.
It can be shown using a recursive technique
(see e.g., Tsypkin \cite{Tsy71},
 Gy\"orfi \cite{Gyo84}, Singer and Feder \cite{SiFe00}, and Gy\"orfi and Lugosi \cite{Laci2})
that the $c_{n,j}$ can be calculated with small computational complexity.

The experts are mixed via an exponential weighting, which is defined the same way as
earlier. Thus, the aggregated prediction scheme is
\begin{equation*}
g_n(x_1^n,y_1^{n-1})=\sum_{k,\ell=1}^{\infty}p_{k,\ell,n}h_n^{(k,\ell)}(x_1^{n},y_1^{n-1}),
\qquad n=1,2, \hdots,
\end{equation*}
where the $p_{k,\ell,n}$ are calculated according to (\ref{eq:pkln}).

Combining the
 proof of Theorem  \ref{th:kernel} and the proof of Theorem 2 in \cite{Laci2} leads to the following result:
\begin{theo}\label{th:gl}
Suppose that $|\phi_j^{(k)}|\le 1$ and, for any
fixed $k$, suppose that the set
\[
\left\{\sum_{j=1}^{\ell}c_j\phi_j^{(k)};\ \ (c_{1},\dots
,c_{\ell}),\ \ell=1,2,\dots\right\}
\]
is dense in the set of continuous functions of $d(k+1)+k$ variables.
Then the generalized linear prediction strategy defined above is universally consistent with respect to the class of
all jointly stationary and ergodic processes such that
\[
\EXP\{Y_0^4\}<\infty.
\]
\end{theo}
We give a sketch of the proof of Theorem \ref{th:gl} in Section \ref{proofs}.
\subsection{Prediction of Gaussian processes}
\label{gaussproc}
We consider in this section the classical problem of Gaussian time series prediction (cf. Brockwell and
Davis \cite{Brockwell}). In this context, parametric models based on
distribution assumptions and structural conditions such
as AR($p$), MA($q$), ARMA($p$,$q$) and ARIMA($p$,$d$,$q$) are usually fitted to
the data (cf. Gerencs\'er and Rissanen \cite{GeRi86}, Gerencs\'er \cite{Ger92, Ger94}, Goldenshluger and Zeevi \cite{GoZe99}). However, in the spirit of modern nonparametric inference, we try to avoid such
restrictions on the process structure. Thus, we only
assume that we observe a string realization $y_1^{n-1}$ of a zero
mean, stationary and ergodic, Gaussian process $\{Y_n\}_{-\infty}^{\infty}$, and try to predict $y_n$, the value of the process at time $n$. Note that there is no side information vectors $x_1^n$ in this purely time series prediction framework.

It is
well known for Gaussian time series that the best predictor
is a linear function of the past:
\[
\EXP\{Y_n\mid Y_{n-1},Y_{n-2},\ldots\}=\sum_{j=1}^{\infty}c_j^* Y_{n-j},
\]
where the $c^*_{j}$ minimize the criterion
\begin{equation*}
\EXP\left\{
\left(\sum_{j=1}^{\infty} c_{j} Y_{n-j}-Y_n\right)^2\right\}.
\end{equation*}

Following Gy\"orfi and Lugosi \cite{Laci2}, we extend the principle of generalized
linear estimates to the prediction of Gaussian time series by considering the special case
$$
\phi_j^{(k)}(y_{n-k}^{n-1})=y_{n-j} \IND_{\{1 \le j \le k\}},
$$
i.e.,
\[
  \tilde{h}_n^{(k)}(y_1^{n-1}) = \sum_{j=1}^{k} c_{n,j}y_{n-j}.
\]
Once again, the coefficients $c_{n,j}$ are calculated according to the past observations $y_{1}^{n-1}$
by minimizing the criterion:
\[
\sum_{t=k+1}^{n-1} \left(\sum_{j=1}^{k}
c_{j}y_{t-j}-y_t\right)^2
\]
if $n>k$, and the all-zero vector otherwise.

With respect to the combination of elementary experts
$\tilde{h}^{(k)}$, Gy\"orfi and Lugosi applied in \cite{Laci2} the so-called ``doubling-trick'', which means that
the time axis is segmented into exponentially increasing
epochs and at the beginning of each epoch the
forecaster is reset.

In this section we propose a much simpler procedure which avoids in particular the
doubling-trick. To begin, we set
\[
  h_n^{(k)}(y_1^{n-1}) = T_{\min\{n^{\delta},k\}}\left(\tilde{h}_n^{(k)}(y_1^{n-1})\right),
\]
where \[
0 <\delta <\frac{1}{8},
\]
and combine these experts as before. Precisely, let $\{q_{k}\}$ be an arbitrarily
probability distribution over the positive integers such that for all $k$, $q_{k}>0$, and for $\eta_n>0$, define the weights
\[
w_{k,n}=q_{k}e^{-\eta_n(n-1)L_{n-1}(h_n^{(k)})}
\]
and their normalized values
\[
p_{k,n}=\frac{w_{k,n}}{\sum_{i=1}^{\infty}w_{i,n}}.
\]
The prediction strategy $g$ at time $n$ is defined by
\begin{equation*}
  g_n(y_1^{n-1})
  = \sum_{k=1}^{\infty}p_{k,n}h_n^{(k)}(y_1^{n-1}),
\qquad n=1,2,\ldots
\end{equation*}
By combining the proof of Theorem \ref{th:kernel} and Theorem 3 in
\cite{Laci2}, we obtain the following result:

\begin{theo}
\label{gauss}
The linear prediction strategy $g$ defined above is universally consistent with respect to the class of all jointly stationary and ergodic
zero-mean Gaussian processes.
\end{theo}

The following corollary shows that the strategy $g$ provides asymptotically a
good estimate of the regression function in the following sense:
\begin{cor}[Gy\"orfi and Ottucs\'ak \cite{Laci4}] \label{cor1} Under the conditions of Theorem \ref{gauss},
\[
\lim_{n\to\infty} \frac{1}{n}\sum_{t=1}^n
\left(\EXP\{Y_t\mid Y_1^{t-1}\}-g(Y_1^{t-1})\right)^2=0 \quad \mbox{almost surely}.
\]
\end{cor}

Corollary \ref{cor1} is expressed in terms of an almost sure Ces\'aro consistency. It is an open problem to know whether there exists a prediction rule $g$ such that
\begin{equation} \label{eq:oq}
\lim_{n\to\infty} \left(\EXP\{Y_n|Y_1^{n-1}\}-g(Y_1^{n-1})\right)=0 \quad \mbox{almost surely}
\end{equation}
for all stationary and ergodic Gaussian processes. Sch\"afer \cite{Sch02}
proved that, under some conditions on the time series, the consistency (\ref{eq:oq}) holds.
\section{Experimental results and analyses}
We evaluated the performance of the histogram,  moving-window kernel, NN and Gaussian process strategies
on two real world data sets. Furthermore, we compared these performances
to those of the
standard ARMA family of methods on the same data sets.
We show in particular
that the four methods presented in this paper usually perform better than
the best ARMA results, with respect to three different criteria.

\begin{figure}[!ph]
\centering
\includegraphics*[width=10truecm, height=7truecm]{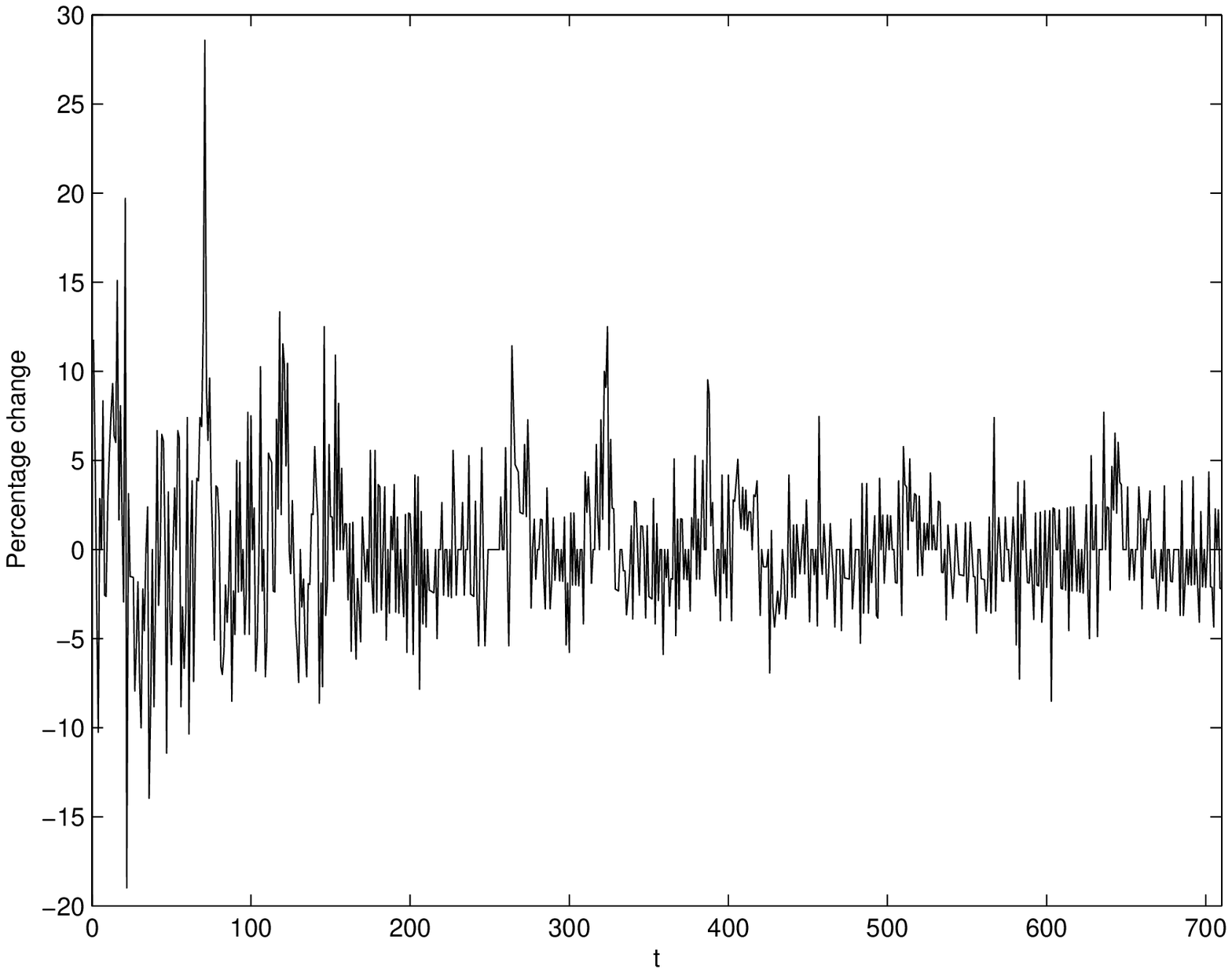}
\caption{{\small Monthly percentage change in USA unemployment rate for January
1948 until March 2007.}}
\label{unem}
\vspace{1cm}
\includegraphics*[width =10truecm, height =7truecm]{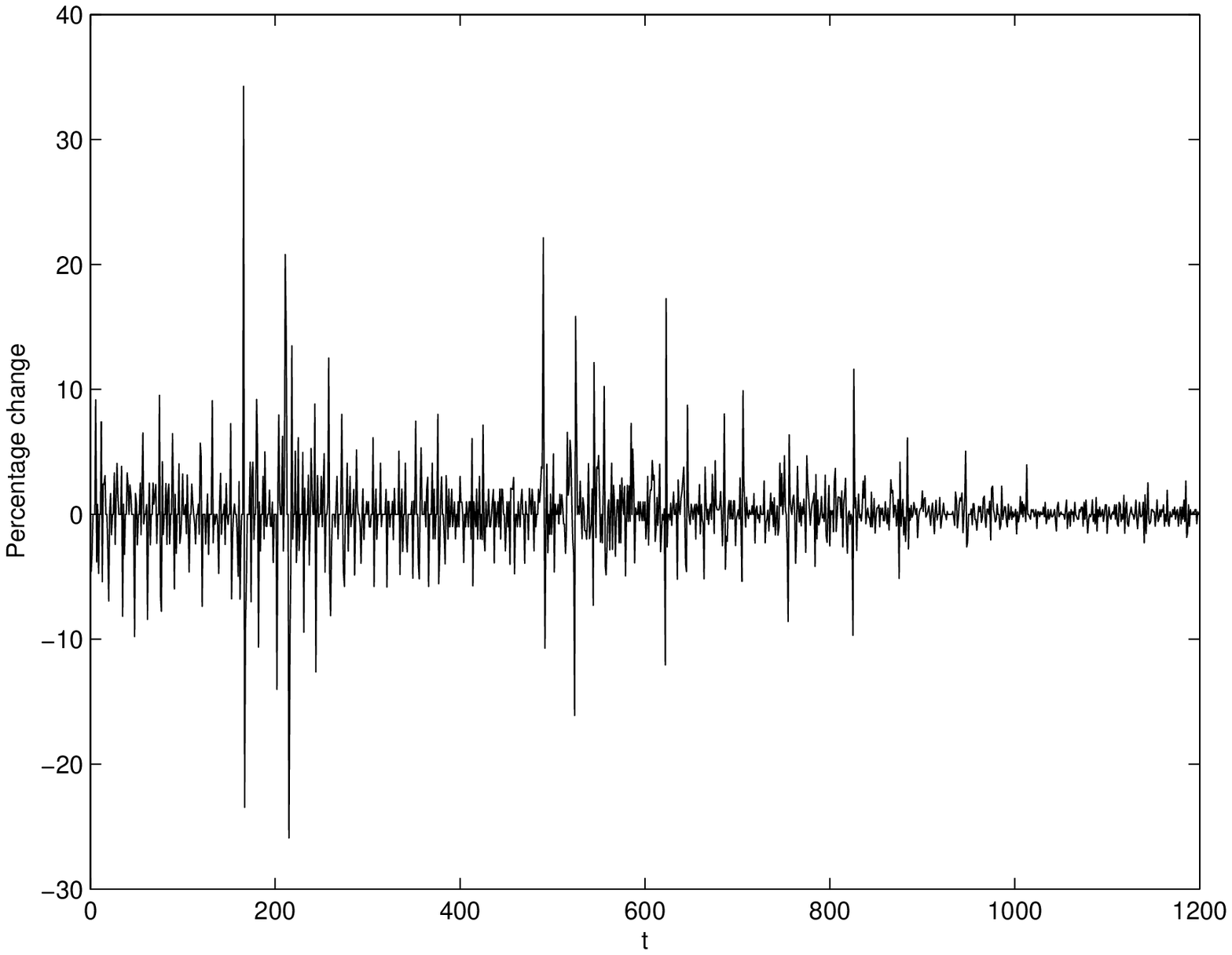}
\caption{{\small Daily percentage change in USA federal funds interest rate for
12 January 2003 until 21 March 2007.}}
\label{time}
\end{figure}

The two real-world time series we investigated were the monthly USA unemployment rate for January
1948 until March 2007 (710 points) and daily USA federal funds interest rate for
12 January 2003 until 21 March 2007 (1200 points) respectively, extracted from the
website {\it economagic.com}. In order to remove first-order trends, we transformed these
time series into time series of {\it percentage change} compared to the previous month or day,
respectively. The resulting time series are shown in Figs. \ref{unem} and \ref{time}.

Before testing the four methods of the present paper alongside the ARMA methods, we tested whether the resulting
time series were trend/level stationary using two standard tests, the KPSS test \cite{kpss} and
the PP test \cite{pp}. For both series using the KPSS test, we did not reject the null hypothesis
of level stationarity at $p = .01, .05$ and $.1$ respectively, and for both series using the PP test
(which has for null hypothesis the existence of a unit root and for alternative hypothesis,
level stationarity), the null hypothesis was rejected at $p= .01, .05$ and $.1$. 

We remark that
this means the ARIMA$(p,d,q)$ family of models, richer than ARMA$(p,q)$
is unnecessary, or equivalently, we need only to consider the ARIMA family ARIMA$(p,0,q)$.
As well as this, the Gaussian process method requires the normality of the data. Since
the original data in both data sets is discretized (and not very finely), this meant that
the data, when transformed into percentage changes only took a small number of fixed values.
This had the consequence that directly applying standard normality tests gave curious results even
when histograms of the data appeared to have near-perfect Gaussian forms; however
adding small amounts of random noise to the data allowed us to
not systematically reject the hypothesis of normality.

Given each method and each time series $(y_1,\ldots,y_m)$ (here, $m=710$ or $1200$), for each $15 \leq n \leq m-1$
we used the data $(y_1,\ldots,y_n)$ to predict the value of $y_{n+1}$. We used three criteria
to measure the quality of the overall set of predictions. First, as described in the present paper,
we calculated the normalized cumulative prediction squared error $L_m$ (since
we start with $n=15$ for practical reasons, this is almost but not exactly what
has been called $L_n$ until now). Secondly, we calculated $L_{m}^{50}$, the normalized cumulative
prediction error over only the last 50 predictions of the time series in order to
see how the method was working after having learned nearly the whole time series. Thirdly,
since in practical situations we may want to predict only the {\it direction} of change, we compared
the direction (positive or negative) of the last $50$ predicted points with respect to each previous, known point,
to the $50$ real directions. This gave us the criteria
$A^{50}$: the
{\it percentage of the direction of the last 50
points correctly predicted}.

\begin{sloppypar}
As in  \cite{Laci3} and \cite{Laci6}, for practical reasons we chose a finite
grid of experts: $k = 1,\ldots,K$ and $\ell = 1,\ldots,L \,$ for the histogram,
kernel and NN
strategies,
fixing $K = 5$ and $L = 10$.
For the histogram strategy we partitioned the space into each of
$\{2^2,\,2^3,\ldots,2^{11}\}$ equally sized intervals, 
for the kernel strategy we
let the radius $r'_{k,\ell}$ take the values
$r'_{k,\ell} \in \{.001,\, .005,\, .01,\, .05,\, .1,\, .5,\, 1,\, 5,\, 10,\, 50\}$ and 
for the NN strategy we set $\bar{\ell} = \ell.$
Furthermore, we fixed the probability distribution $\{q_{k,\ell}\}$
as the uniform distribution over the $K \times L$ experts. For the Gaussian process method,
we simply let $K=5$ and fixed the probability distribution $\{q_{k}\}$
as the uniform distribution over the $K$ experts.
\end{sloppypar}

Used to compare standard methods with the present nonparametric str\-a\-t\-egies, the ARMA$(p,q)$ algorithm was run
for all pairs $(p,q) \in \{0, 1, 2, 3, 4, 5\}^2$. The ARMA family of methods is a combination
of an autoregressive part AR$(p)$ and a moving-average part MA$(q)$.
Tables \ref{unemp} and \ref{timetest} show the histogram, kernel, NN, Gaussian process
and ARMA results for the
unemployment and interest rate time series respectively.
The three ARMA results shown in each table are those which
had the best $L_m$, $L_{m}^{50}$ and $A^{50}$ respectively (sometimes two or more had
the same $A^{50}$, in which case we chose one of these randomly). The best results with respect
to each of the three criteria are
shown in bold.

\begin{table}[!hf]

    \centering
        \begin{tabular}{|c||c|c|c|}
            \hline
            & $L_m$   & $L_{m}^{50}$ & $A^{50}$  \\
            \hline \hline
             histogram  & 15.66 & \textbf{4.82} & 68 \\
            \hline
             kernel  & 15.44 & 4.99 & 68 \\
            \hline
            NN    & \textbf{15.40} &  4.97  &   70 \\
            \hline
            Gaussian    & 16.35 &  5.02  &   76 \\
            \hline
            ARMA$(1,1)$  & 16.26 & 5.31 & 72  \\
            \hline
            ARMA$(0,0)$ &  16.68 &  4.86 & \textbf{78} \\
            \hline
            ARMA$(2,0)$ & 16.46 & 5.12 & \textbf{78}\\
            \hline

        \end{tabular}

\caption{{\small Results for histogram, kernel, NN, Gaussian process and ARMA prediction methods
on the monthly percentage change in USA unemployment rate from January
1948 until March 2007. The three ARMA results are those which
performed the best in terms of the  $L_m$, $L_{m}^{50}$ and $A^{50}$ criteria
respectively}.  \label{unemp}}
\end{table}


\begin{table}[!hf]

    \centering
        \begin{tabular}{|c||c|c|c|}
            \hline
            & $L_m$   & $L_{m}^{50}$ & $A^{50}$  \\
            \hline \hline
             histogram  & 9.78 & \textbf{0.52} & \textbf{88} \\
            \hline
             kernel  & \textbf{9.77} & 0.57 & 86 \\
            \hline
            NN    & 9.86 &  0.79  &   80 \\
            \hline
            Gaussian    & 9.98 &  0.62  &   82 \\
            \hline
            ARMA$(1,1)$  & 9.90 & 0.78 & 70  \\
            \hline
            ARMA$(0,1)$ &  10.30 &  0.60 & 82 \\
            \hline
            ARMA$(3,0)$ & 10.12 & 0.63 & \textbf{88}\\
            \hline

        \end{tabular}

\caption{{\small Results for histogram, kernel, NN, Gaussian process and ARMA prediction methods
on the daily percentage change in the USA federal funds interest rate from
12 January 2003 until 21 March 2007. The three ARMA results are those which
performed the best in terms of the  $L_m$, $L_{m}^{50}$ and $A^{50}$ criteria
respectively.} \label{timetest}}
\end{table}

We see via Tables \ref{unemp} and \ref{timetest}
that the histogram, kernel and NN strategies presented here outperform all 36 possible ARMA$(p,q)$ models
($0 \leq p,q \leq 5$) in terms of normalized cumulative prediction error $L_m$, and that the
Gaussian process method performs similarly to the best ARMA method.
In terms of the $L_{m}^{50}$ and $A^{50}$ criteria, all of the present
methods and the best ARMA method provide broadly similar results.
From a practical point of view, we note also that the histogram, kernel and NN methods also run much faster than a single
ARMA$(p,q)$ trial on a standard desktop computer. For example, the NN method is of the order of 10 to 100 times
faster than an ARMA$(p,q)$ for a time series with about 1000 points, depending on the values
of $p$ and $q$.

\section{Proofs} \label{proofs}

\subsection{Proof of Theorem \ref{th:kernel}}
The proof of Theorem \ref{th:kernel} strongly relies on the  following two lemmas. The first one is known as Breiman's generalized
ergodic theorem. 
\begin{lem}[Breiman \cite{Breiman}]
\label{Breiman} Let $Z=\{Z_n\}_{-\infty}^{\infty}$ be a stationary and
ergodic process. For each positive integer $t$, let $T^t$ denote the
left shift operator, shifting any sequence $\{\hdots, z_{-1},
z_0,z_1, \hdots\}$ by $t$ digits to the left. Let $\{f_t\}_{t \geq 1}$
be a sequence of real-valued functions such that $\lim_{t \to
\infty}f_t(Z) =f(Z)$ almost surely for some function $f$. Suppose
that $\mathbb E\sup_t |f_t(Z)|<\infty$. Then
$$\lim_{n \to \infty} \frac{1}{n} \sum_{t=1}^n f_t(T^tZ)=\mathbb E \left\{f(Z)\right\}\quad \mbox{almost surely}.$$
\end{lem}

\begin{lem}[Gy\"orfi and Ottucs\'ak \cite{Laci4}]
\label{expert} Let $h^{(1)},h^{(2)},\dots$ be a sequence of
prediction strategies (experts). Let $\{q_{k}\}$ be a probability
distribution on the set of positive integers. Denote the
normalized loss of any expert $h=\{h_n\}_{n=1}^{\infty}$ by
\[
L_n(h) = \frac{1}{n}\sum_{t=1}^n \mathcal L(h_t,Y_t),
\]
where the loss function $\mathcal L$ is convex in its first argument $h_t$.
Define
\[
w_{k,n}=q_{k}e^{-\eta_n(n-1)L_{n-1}(h^{(k)})},
\]
where $\eta_n > 0$ is monotonically decreasing, and set
\[
p_{k,n}= \frac{w_{k,n}}{\sum_{i=1}^{\infty}w_{i,n}}.
\]
If the prediction strategy $g=\{g_n\}_{n=1}^{\infty}$ is defined by
\[
  g_n = \sum_{k=1}^{\infty} p_{k,n}h^{(k)}_n,
\qquad n=1,2,\ldots
\]
then, for every $n \ge 1$,
\[
L_n(g) \le \inf_{k}\lt( L_n(h^{(k)})-\frac{\ln
q_{k}}{n\eta_{n+1}}\rt) +\frac{1}{2n} \sum_{t=1}^n
\eta_t \sum_{k=1}^\infty p_{k,t} \mathcal L^2(h_t^{(k)},Y_t).
\]
\end{lem}

\noindent\textbf{Proof of Theorem \ref{th:kernel}.}
Because of (\ref{algoet}) it is enough to show that
\[
\limsup_{n\to \infty}L_n(g)\le L^* \quad \mbox{almost surely}.
\]
With this in mind, we introduce the following notation:
\[
\widehat{E}_n^{(k,\ell)}(X_1^{n},Y_1^{n-1},\mathbf{z},\mathbf{s})
=
\frac{\sum_{\left\{k<t<n:\
\|x^{t}_{t-k}-\mathbf{z}\| \le
r_{k,\ell},\ \|y^{t-1}_{t-k}-\mathbf{s}\|\le r'_{k,\ell}
\right\}} Y_t} {\left|
\left\{k<t<n:\
\|x^{t}_{t-k}-\mathbf{z}\| \le
r_{k,\ell},\ \|y^{t-1}_{t-k}-\mathbf{s}\|\le r'_{k,\ell}
\right\}\right|}
\]
for all $n>k+1$, where $0/0$ is defined to be $0$,
$\mathbf{z} \in (\R^{d})^{k+1}$ and $\mathbf{s} \in \R^{k}$. Thus, for any $h^{(k,\ell)}$, we can write
\[
h_n^{(k,\ell)}(X_1^{n},Y_1^{n-1})=T_{\min{\{n^{\delta},\ell\}}} \left(\widehat{E}_n^{(k,\ell)}(X_1^{n},Y_1^{n-1},X_{n-k}^{n},Y_{n-k}^{n-1})\right).
\]

By a double application of the ergodic theorem, as $n \to \infty$, almost
surely, for a fixed $\mathbf{z} \in (\mathbb{R}^{d})^{k+1}$ and
$\mathbf{s} \in \mathbb{R}^{k}$, we may write
\begin{align*}
\widehat{E}_n^{(k,\ell)}(X_1^{n},Y_1^{n-1},\mathbf{z},\mathbf{s})&
= \frac{\frac{1}{n}\sum_{\left\{k<t<n:\
\|X^{t}_{t-k}-\mathbf{z}\| \le
r_{k,\ell},\ \|Y^{t-1}_{t-k}-\mathbf{s}\|\le r'_{k,\ell}
\right\}} Y_t} {\frac{1}{n} \left|
\left\{k<t<n:\
\|X^{t}_{t-k}-\mathbf{z}\| \le
r_{k,\ell},\ \|Y^{t-1}_{t-k}-\mathbf{s}\|\le r'_{k,\ell}
\right\}\right|}\\
&\to \frac{\EXP \{Y_0\IND_{\{\|X^{0}_{-k}-\bz\| \le
r_{k,\ell},\ \|Y^{-1}_{-k}-\bs\|\le r'_{k,\ell}\}}\}}
{\mathbb P \left\{\|X^{0}_{-k}-\bz\| \le
r_{k,\ell},\ \|Y^{-1}_{-k}-\bs\|\le r'_{k,\ell}\right\}}\\
& = \EXP \{Y_0\,|\,\|X^{0}_{-k}-\bz\| \le
r_{k,\ell},\ \|Y^{-1}_{-k}-\bs\|\le r'_{k,\ell}\}.
\end{align*}
Therefore, for all $\bz$ and $\bs$,
\begin{eqnarray*}
\lefteqn{\lim_{n \to \infty} T_{\min{\{n^{\delta},\ell\}}} \left(\widehat{E}_n^{(k,\ell)}(X_1^{n},Y_1^{n-1},\mathbf{z},\mathbf{s})\right)}\\
&=& T_{\ell}\left( \EXP \{Y_0\,|\,\|X^{0}_{-k}-\bz\| \le r_{k,\ell},\ \|Y^{-1}_{-k}-\bs\|\le r'_{k,\ell}\}\right) \\
&\defeq&  \varphi_{k,\ell}(\bz,\bs).
\end{eqnarray*}
Thus, by Lemma \ref{Breiman}, as $n\to \infty$, almost surely,
\begin{align}
L_n(h^{(k,\ell)}) & =\frac{1}{n}\sum_{t=1}^n (h_t^{(k,\ell)}(X_1^t,Y_1^{t-1})-Y_t)^2 \notag \\*
&  =\frac{1}{n}\sum_{t=1}^n \left(
T_{\min{\{t^{\delta},\ell\}}} \left(\widehat{E}_t^{(k,\ell)}(X_1^{t},Y_1^{t-1},X_{t-k}^{t},Y_{t-k}^{t-1})\right)-Y_t\right)^2 \notag\\*
& \to \EXP \left\{(\varphi_{k,\ell}(X_{-k}^{0},Y_{-k}^{-1})-Y_0)^2\right\} \notag \\* 
& \defeq\varepsilon_{k,\ell}~\notag.
\end{align}

Denote, for Borel sets $A \subset (\mathbb R^{d})^{k+1}$ and $B \subset \mathbb R^k$,
$$ \mu_k(A,B)\defeq\mathbb P\{X_{-k}^{0} \in A, Y_{-k}^{-1} \in B\},$$
and set
$$\psi_k(\bz,\bs)\defeq\mathbb E\{Y_0\,|\,X_{-k}^{0}=\bz, Y_{-k}^{-1}=\bs\}.$$
Next, let $S_{\bs,r}$ denote the closed ball with center  $\bs$ and radius $r$. 
Let
$$\tilde \varphi_{k,\ell}(\mathbf z,\mathbf s) \defeq \EXP \{Y_0\,|\,\|X^{0}_{-k}-\bz\| \le r_{k,\ell},\ \|Y^{-1}_{-k}-\bs\|\le r'_{k,\ell}\},$$
then for any $\bz$ and $\bs$  which are in the support  of $\mu_k$, we have
\begin{align*}
\varphi_{k,\ell}(\bz,\bs)& = T_\ell \left(\EXP \{Y_0\,|\,\|X^{0}_{-k}-\bz\| \le
r_{k,\ell},\ \|Y^{-1}_{-k}-\bs\|\le r'_{k,\ell}\}\right)\\
&=
T_{\ell}\left(\frac{\EXP \{Y_0\IND_{\{\|X^{0}_{-k}-\bz\| \le
r_{k,\ell},\ \|Y^{-1}_{-k}-\bs\|\le r'_{k,\ell}\}}\}}
{\mathbb P \left\{\|X^{0}_{-k}-\bz\| \le
r_{k,\ell},\ \|Y^{-1}_{-k}-\bs\|\le r'_{k,\ell}\right\}}\right)\\
& = T_{\ell}\left(\frac{1}{\mu_k(S_{\bz,r_{k,\ell}},S_{\bs,r'_{k,\ell}})}
\int_{x \in S_{\bz,r_{k,\ell}},~y \in  S_{\bs,r'_{k,\ell}}}\tilde \varphi_{k,\ell}(x,y)\,\mu_k(\mbox{d}x,\mbox{d}y)\right)\\
&  \to \psi_{k}(\bz,\bs),
\end{align*}
as $\ell \to \infty$ and for $\mu_k$-almost all $\bs$ and $\bz$ by the Lebesgue density theorem (see Gy\"orfi, Kohler, Krzy\.zak and Walk \cite{Laciregressionbook}, Lemma 24.5). Therefore,
\begin{equation*}
\label{demenagement1}
\lim_{\ell \to \infty}\varphi_{k,\ell}(X_{-k}^{0},Y_{-k}^{-1})
=\psi_k(X_{-k}^{0},Y_{-k}^{-1})\quad \mbox{almost surely}.
\end{equation*}
Observe that
\begin{align*}
\varphi_{k,\ell}^2(\bz,\bs)&=\left[T_{\ell}\left(\mathbb E\{Y_0\,|\,\|X^{0}_{-k}-\bz\| \le
r_{k,\ell},\ \|Y^{-1}_{-k}-\bs\|\le r'_{k,\ell}\}\right)\right]^2\\
& \leq \left(\mathbb E\{Y_0\,|\,\|X^{0}_{-k}-\bz\| \le
r_{k,\ell},\ \|Y^{-1}_{-k}-\bs\|\le r'_{k,\ell}\}\right)^2\\
& \qquad \mbox{(since $|T_{\ell}(z)| \leq |z|$)}\\
& \leq \mathbb E \{Y_0^2\,|\, \|X^{0}_{-k}-\bz\| \le
r_{k,\ell},\ \|Y^{-1}_{-k}-\bs\|\le r'_{k,\ell}\}\\
& \qquad \mbox{(by Jensen's inequality).}\\
\end{align*}
Consequently,
\begin{equation*}
\label{demenagement2}
\sup_{\ell \geq 1}\mathbb E\{ \varphi_{k,\ell}^2(X_{-k}^{0},Y_{-k}^{-1})\}\leq \mathbb E Y_0^2 <\infty,
\end{equation*}
due to the assumptions of the theorem.
Therefore, for fixed $k$ the sequence of random variables $\{\varphi_{k,\ell}(X_{-k}^{0},Y_{-k}^{-1})\}_{\ell=1}^{\infty}$ is uniformly integrable and by
using the dominated convergence theorem we obtain
\begin{align*}
\lim_{\ell \to \infty}\varepsilon_{k,\ell} & = \lim_{\ell \to \infty}\EXP \left\{(\varphi_{k,\ell}(X_{-k}^{0},Y_{-k}^{-1})-Y_0)^2\right\}\\
& =\mathbb E\left\{\left(\mathbb E\{Y_0|X_{-k}^{0},Y_{-k}^{-1}\}-Y_0\right)^2\right\}\\
& \defeq\varepsilon_{k}.
\end{align*}
Invoking the martingale convergence theorem  (see, e.g., Stout \cite{Sto74}), we then have
$$\lim_{k\to \infty} \varepsilon_k=\mathbb E\left\{\left(\mathbb E\{Y_0|X_{-\infty}^{0},Y_{-\infty}^{-1}\}-Y_0\right)^2\right\}=L^*,$$
and consequently,
$$\lim_{k,\ell \to \infty} \varepsilon_{k,\ell}=L^*.$$

We next apply Lemma \ref{expert} with the choice
$\eta_n=1/\sqrt{n}$ and the squared loss
$$\mathcal L(h_t,Y_t)=(h_t-Y_t)^2.$$
We obtain
\begin{align*}
L_n(g) & \le \inf_{k,\ell}\lt( L_n(h^{(k,\ell)})-\frac{2\ln
q_{k,\ell}}{\sqrt{n}}\rt) \\
& \quad +\frac{1}{2n} \sum_{t=1}^n
\frac{1}{\sqrt{t}} \sum_{k,\ell=1}^\infty p_{k,\ell,t} \left(h_t^{(k,\ell)}(X_1^t,Y_1^{t-1})-Y_t\right)^4.
\end{align*}
On one hand, almost surely,
\begin{align*}
& \limsup_{n\to \infty} \inf_{k,\ell}
\lt(L_n(h^{(k,\ell)}) - \frac{2\ln q_{k,\ell}}{\sqrt{n}} \rt)\\
& \quad \le \inf_{k,\ell} \limsup_{n\to \infty}
\lt(L_n(h^{(k,\ell)}) - \frac{2\ln q_{k,\ell}}{\sqrt{n}} \rt)\\
& \quad =  \inf_{k,\ell} \limsup_{n\to \infty}
L_n(h^{(k,\ell)}) \\
&\quad  =  \inf_{k,\ell}\varepsilon_{k,\ell}\\
& \quad \leq  \lim_{k,\ell\to \infty}\varepsilon_{k,\ell}\\
& \quad = L^*.
\end{align*}
On the other hand,
\begin{align*}
& \frac{1}{n} \sum_{t=1}^n \frac{1}{\sqrt{t}}\sum_{k,\ell=1}^{\infty} p_{k,\ell,t} (h_t^{(k,\ell)}(X_1^t,Y_1^{t-1})-Y_t)^4\\
& \quad \le \frac{8}{n} \sum_{t=1}^n \frac{1}{\sqrt{t}}
\sum_{k,\ell=1}^{\infty} p_{k,\ell,t} \lt(h_t^{(k,\ell)}(X_1^t,Y_1^{t-1})^4 +Y_t^4\rt)\\
& \quad \le \frac{8}{n} \sum_{t=1}^n \frac{1}{\sqrt{t}}\sum_{k=1}^{\infty}
\lt(\sum_{\ell=1}^{\lfloor t^{\delta}\rfloor} p_{k,\ell,t} \ell^4 +
\sum_{\ell=\lceil t^{\delta}\rceil}^{\infty} p_{k,\ell,t}t^{4\delta} +
\sum_{\ell=1}^{\infty}p_{k,\ell,t}Y_t^4\rt) \\
& \quad \le \frac{8}{n} \sum_{t=1}^n \frac{1}{\sqrt{t}}\sum_{k,\ell=1}^{\infty} p_{k,\ell,t} \lt(t^{4\delta} + Y_t^4\rt)\\
& \quad =\frac{8}{n} \sum_{t=1}^n \frac{t^{4\delta} + Y_t^4}{\sqrt{t}}.
\end{align*}
Therefore, almost surely,
\begin{align*}
& \limsup_{n\to \infty} \frac{1}{n} \sum_{t=1}^n \frac{1}{\sqrt{t}}
\sum_{k,\ell=1}^{\infty} p_{k,\ell,t} (h_t^{(k,\ell)}(X_1^t,Y_1^{t-1})- Y_t)^4\\*
&\quad \le\quad\! \limsup_{n\to \infty} \frac{8}{n} \sum_{t=1}^n
\frac{Y_t^4}{\sqrt{t}}\\
&\quad =0\\
& \qquad \mbox{(since $\delta<1/8$ and $\EXP\{Y_0^4\}<\infty $).}
\end{align*}
Summarizing these bounds, we get that, almost surely,
\[
\limsup_{n\to \infty} L_n(g) \le L^*,
\]
and the theorem is proved. \qed

\subsection{Sketch of the proof of Theorem \ref{th:gl}}
For fixed $k$ and $\ell$, let
\begin{equation*}
(c^*_{1},\dots ,c^*_{\ell}) \in \arg\min_{(c_{1},\dots ,c_{\ell})}
\EXP\left\{
\left(\sum_{j=1}^{\ell} c_{j}\phi_j^{(k)}(X_{-k}^{0},Y_{-k}^{-1})-Y_0\right)^2\right\}.
\end{equation*}
Then, following the proof of Theorem 2 in \cite{Laci2} one can show
that for all $j \in \{1,\ldots, \ell\}$,
\begin{equation}
\lim_{n\to \infty} c_{n,j}=c^*_j \quad \mbox{almost surely}, \label{cc}
\end{equation}
where the $c_{n,j}$ are defined in (\ref{eq:cnj}). Using equality (\ref{cc}) and Lemma \ref{Breiman}, for any fixed $k$ and $\ell$ we obtain that, almost surely,
\begin{eqnarray*}
\lim_{n\to \infty} L_n(h^{(k,\ell )})
& = & \lim_{n\to \infty}
{1\over n}\sum_{t=k+1}^n
\left(h_t^{(k,\ell )}(X_{1}^{t},Y_{1}^{t-1})-Y_t\right)^2\\
&=& \lim_{n\to \infty}{1\over n}\sum_{t=k+1}^n
\left(T_{\min{\{t^{\delta},\ell\}}}\left(\sum_{j=1}^{\ell} c_{t,j}\phi_j^{(k)}(X_{t-k}^{t},Y_{t-k}^{t-1})\right)-Y_t\right)^2\\
&=&\EXP\left\{
\left(T_{\ell}\left(\sum_{j=1}^{\ell} c^*_{j}\phi_j^{(k)}(X_{-k}^{0},Y_{-k}^{-1})\right)-Y_0\right)^2\right\}\\
& \defeq & \varepsilon_{k,\ell}.
\end{eqnarray*}
Then, with similar arguments to Theorem 2 in \cite{Laci2}, it can be shown that
\[
\lim_{k,\ell\to \infty}\varepsilon_{k,\ell} \le L^*.
\]
Finally, by using Lemma \ref{expert}, the assumptions $\delta< 1/8$ and $\EXP \{Y_0^4\} < \infty$, and repeating the arguments of the proof of Theorem \ref{th:kernel}, we obtain
\begin{eqnarray*}
\limsup_{n\to \infty} L_n(g)
 \le  \inf_{k,\ell}\varepsilon_{k,\ell}
 \le  L^*,
\end{eqnarray*}
as desired.
\qed

\end{document}